\documentclass[11pt,twocolumn,twoside,a4paper,amsmath,amssymb,aps,showkeys,showpacs]{revtex4}

\usepackage{amsfonts}
\usepackage{graphics} 
\usepackage{epsfig}
\usepackage{fancyheadings}

\begin{document}
\thispagestyle{myheadings}
\rhead[]{}
\lhead[]{}
\chead[Kokoulina E., Nikitin V., Petukhov Yu., Karpov A. and Kutov A.]
{New results of the extreme multiplicity studies}

\title{New Results of Extreme Multiplicity Studies}

\author{Kokoulina E.}
\altaffiliation[Also at ]{Gomel State Technical University, Gomel,
Belarus} \email{kokoulin@sunse.jinr.ru}
\author{Nikitin V.}
\altaffiliation[]{} \email{nikitin@jinr.sunse.ru} \affiliation{LHEP,
JINR, Dubna, Moscow region, Russia}
\author{Petukhov Yu.}
\altaffiliation[]{} \email{Yuri.Petukhov@ihep.ru} \affiliation{LHEP,
JINR, Dubna, Moscow region, Russia}
\author{Karpov A.}
\email{karpov@komisc.ru}
\author{Kutov A.}
\altaffiliation[]{} \email{kutov@komisc.ru}
\affiliation{Department of Mathematics Komi SC UrD RAS, Syktyvkar,
Russia}
\received{ 23.10.09 }

\begin{abstract}
Extreme multiplicity studies at 50 GeV in pp interactions are
discussed. Preliminary multiplicity distributions at U-70 (IHEP,
Protvino) energy have been obtained for more than 20 charged
particles. A new elaborated algorithm for the track reconstruction
in a drift tube tracker and magnetic spectrometer, has been checked
. The collective behavior of secondary particles is manifested in
these interactions in the extreme multiplicity region. For the first
time the ring events in pp interactions have been observed in this
region. A possibility of detecting the Bose-Einstein condensation
detection is discussed.
\end{abstract}



\maketitle

\renewcommand{\thefootnote}{\fnsymbol{footnote}}

\renewcommand{\thefootnote}{\roman{footnote}}


\section{Introduction}
\label{introduction}

At present the phase transition signal search for hadrons and nuclei
into quark-gluon plasma \cite{Andron,Cleymans,Becattini} and back to
hadronization, is in progress. These studies are carried out at high
and low energies \cite{NuXu}. The data obtained at working
accelerators are checked whether these transitions are possible. The
search for the phase transitions is closely connected with a high
(extreme) multiplicity region exceeding considerably the mean
multiplicity \cite{Therm}.

We suggest that the transition to the quark-gluon phase happens in
this region, and this phase can be revealed by means of the
collective behavior of secondary particles in the proton and nuclear
interactions at accelerator U-70 (IHEP, Protvino). The search is
realized in the extreme multiplicity region because the
Bose-Einstein condensation has been predicted in this very region
\cite{Gorenstein}. Also the indications were obtained on the
formation of ring events (analogy of Cherenkov emission gluons from
quarks) \cite{YaF1} and on the grouping of secondary particles --
clusterization \cite{BJP}.

Now the experimental and theoretical studies are in progress to
check the assumption about the increased soft photon (less than 50
MeV) yield in comparison with the estimations obtained in quantum
electrodynamics. The existing theoretical models and Monte-Carlo
event generators differ very much in their estimations and
multiplicity behavior predictions in the high multiplicity region.

We have developed a Gluon Dominance Model (GDM) \cite{GDM} to
describe the extreme multiplicity and mechanism of hadronization.
This model predicts the maximal number of charged and neutral
particles at 70 GeV which does not exceed 26 and 16,
correspondingly.
\begin{figure}
\includegraphics[width=3.5 in, height=3.2 in, angle=0]{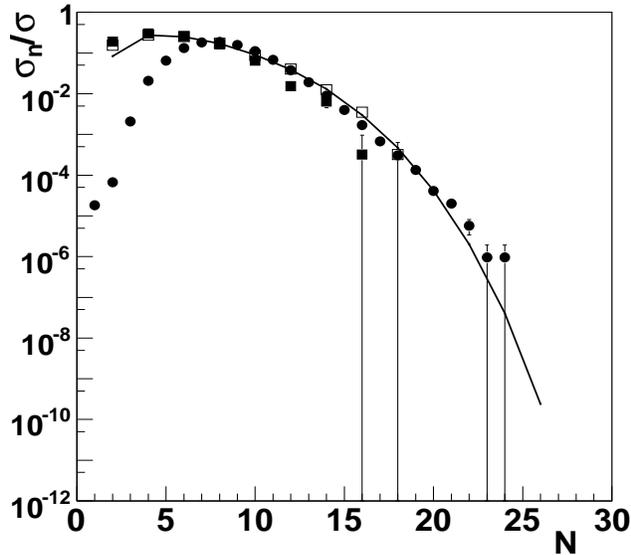}
\caption{\label{1} Multiplicity Distributions obtained at Mirabel:
$\blacksquare $ - $E_p = 50$ GeV and $\Box $ - $E_p = 70$ GeV, at
SVD-2: $\bullet $ - $E_p = 50$ GeV, the solid curve - GDM
prediction.}
\end{figure}

The investigations are carried out on the modern experimental setup
SVD-2 \cite{NPCS} (Spectrometer with Vertex Detector) equipped with
a strip silicon detector, a drift tube tracker, a magnetic
spectrometer with proportional chambers, Cherenkov counter and an
electromagnetic calorimeter.We have designed and manufactured a
scinillator hodoscope (high multiplicity trigger) producing a signal
to register events with multiplicity not less than the given level
-- the so called trigger level. The realization of this project will
enable the community to move ahead towards the deeper understanding
of the strong interaction nature in the extreme multiplicity region.
\section{The SVD-2 setup and data processing}
Since 2005 we have been working with the 50 GeV proton beam. Our
experimental studies are carried out at SVD-2 (Spectrometer with
Vertex Detector) setup \cite{YaF1} on U-70 accelerator of IHEP
(Protvino, Russia).

This installation consists of the following basic elements: a
hydrogen or nuclear target, a precise vertex detector (PVD), a straw
tube chamber or a drift tube tracker (DT), a magnetic spectrometer
(MS) with proportional chambers (PC), Cherenkov counter and an
electromagnetic calorimeter or gamma-quantum detector (DeGa). We
manufactured a scintillation hodoscope or high multiplicity trigger
which produces a signal to record the events with not lower than the
specified multiplicity level.

The main element of SVD-2 setup is PVD. It allows one to reconstruct
the interaction vertex with a high degree of accuracy. It was
manufactured on the basis of strip silicon sensors with a step of 25
and 50 $\mu $m. It has a set of planes at the following angles: $0$,
$\pi /2 $ and $\pm $10.5$^{\circ }$. We distinguish coordinates $x$,
$y$, $u$ (+10.5$^{\circ }$) and $v$ (-10.5$^{\circ }$),
respectively. The oblique planes $U$ and  $V$ necessary for
disentangling of tracks in space, were installed in November 2008.
In the previous 2006 and 2007 runs we obtained double multiplicity
distributions on each from two projections: \emph{XOZ} and
\emph{YOZ}, where axis \emph{Z} is the beam direction.

After the 2008 run we performed data processing  with two oblique
planes and got preliminary single multiplicity distributions using
the PVD data (high multiplicity trigger level was equal to 8) and
compared it with Mirabel data and GDM. Fig. 1 illustrates four
multiplicity distributions for comparison: Mirabel results at 70 GeV
(empty square) and 50 GeV (full square), the last distribution is
consistent with SVD distribution (full circle) and with the gluon
dominance model (the solid curve).

The suppression of distributions in the small multiplicity region (n
$\leq $ 8) is stipulated by the scintillation hodoscope rejections.
This element regulates the high multiplicity event selection. We
have taken into account the efficiency of registration of the high
multiplicity events by the PVD.

The Monte Carlo simulation has demonstrated that the subsequent data
processing from DT and MS does not significantly change these
distributions. These detectors can improve, in principle, the
registration efficiency for high multiplicity events.

The Monte-Carlo simulation and comparison with the Mirabel data have
given the estimations of these losses caused by the limited PVD
acceptance as about two charged particles on the average. These
losses have been taken into account in Fig. 1.

Now we have designed and debugged new software for track
reconstruction by using these additional detectors based on Kalman
Filter technique. This algorithm can use the tracks reconstructed in
PVD and propagate them in DT and MS for a more precise track
parameter estimation (especially momentum) and then it also finds
additional tracks in DT and MS with the hits which were not used in
the previous step. It means that we can find tracks which are
invisible at PVD (because of the restricted acceptance) and
determine the momentum of these particles.

It is also possible to work without PVD information for track
reconstruction but only in DT and MS. The examples of the track
reconstructed events at this energy by means of this algorithm at
three modules of DT and sixteen PC, are shown in Figs. 2 and 3.
\begin{figure}
\includegraphics[width=3.3 in, height=2.7 in, angle=0]{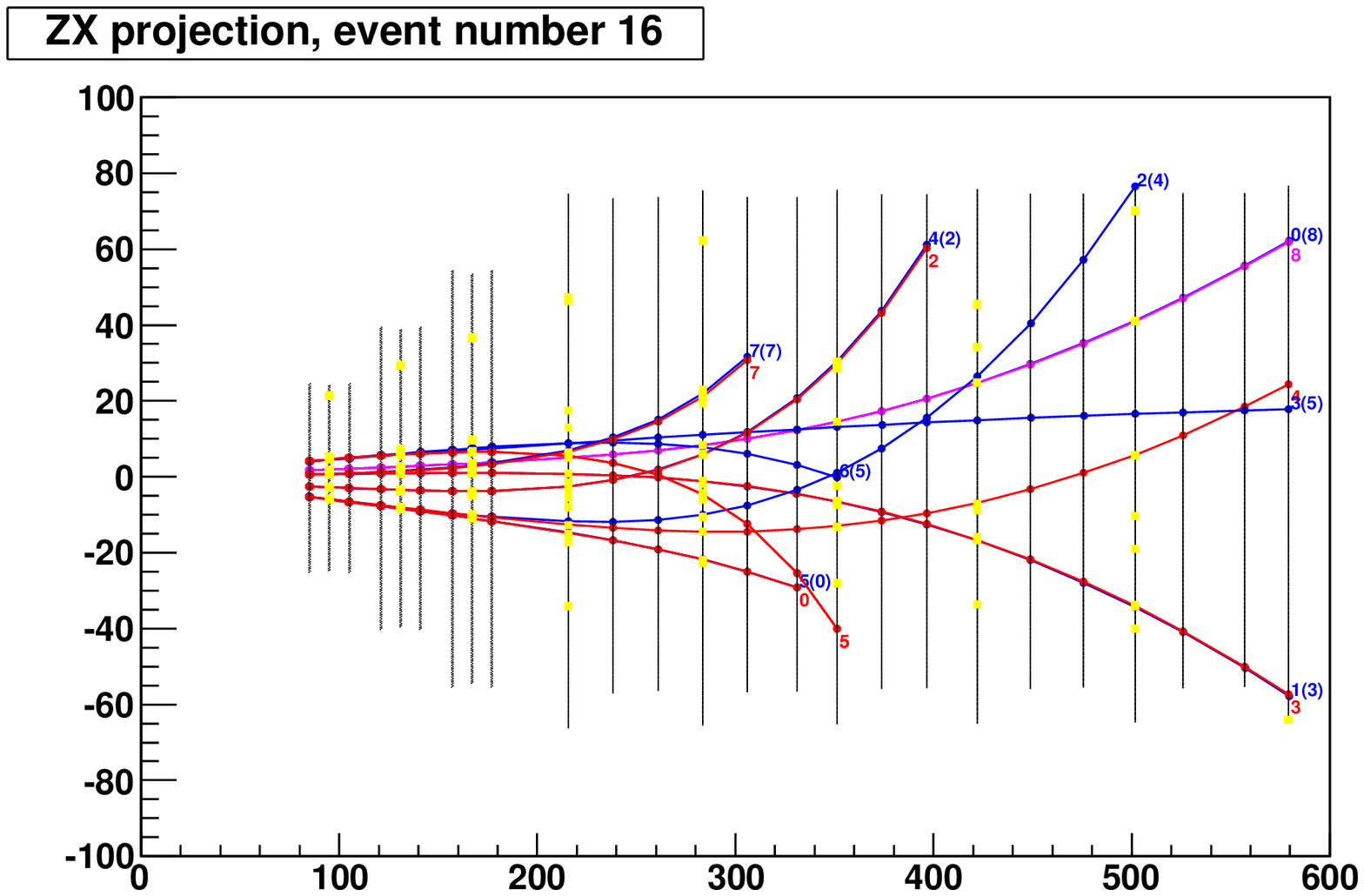}
\caption{\label{2}Example of MC simulated event and its
reconstructed tracks at DT and MS by using a new algorithm.}
\end{figure}
\begin{figure}
\includegraphics[width=3.3 in, height=2.7 in, angle=0]{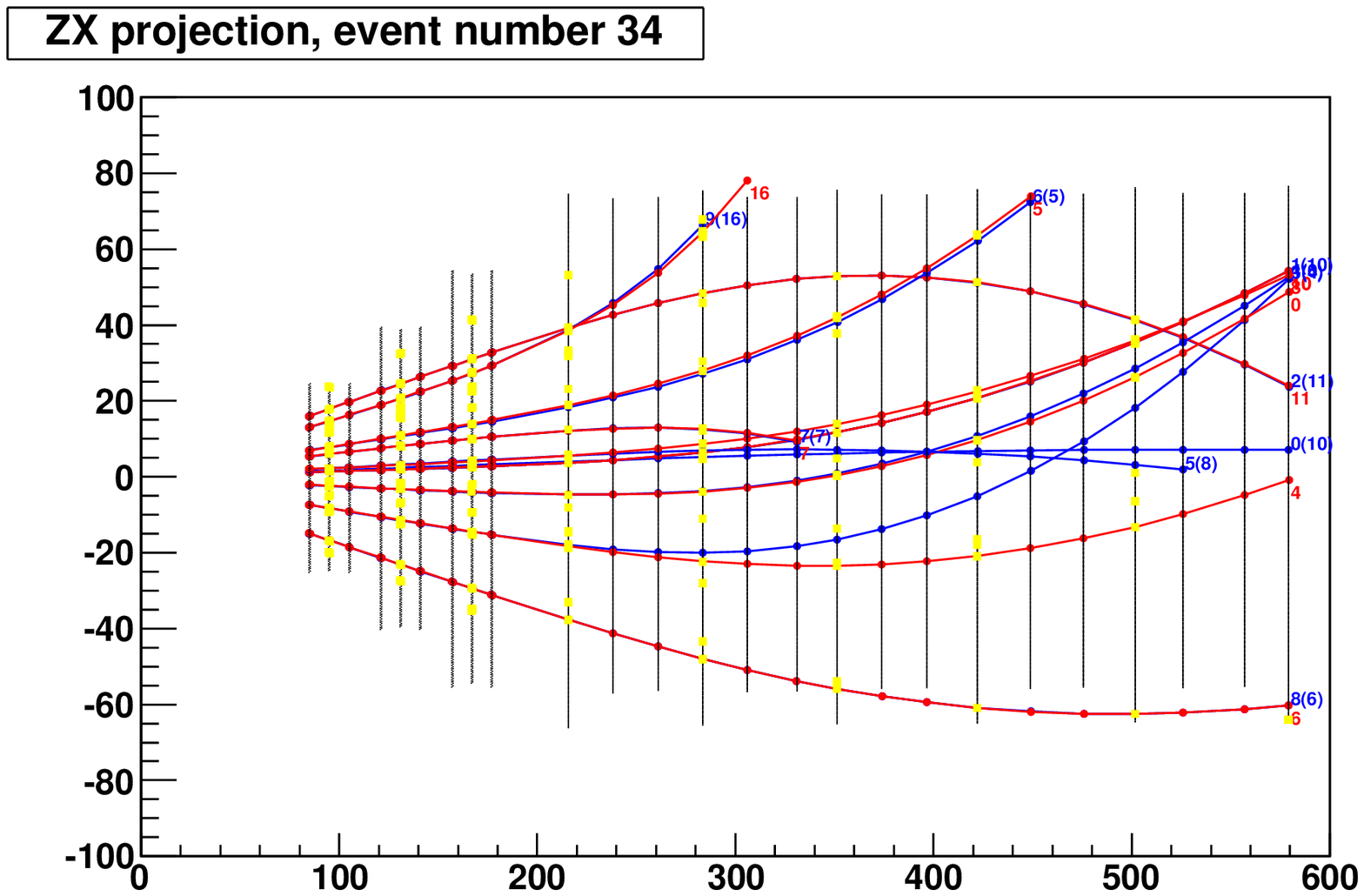}
\caption{\label{3}Example of MC simulated event and its
reconstructed tracks at DT and MS by using a new algorithm.}
\end{figure}
\section{The collective phenomena}
The search for the collective phenomena in the high multiplicity
region is in progress. We consider that they can be observed in this
region. The indications on the ring events were obtained in pA
interactions at high (more than 18) multiplicity \cite{BJP}. It is
interesting to analyze pp interactions and compare them with nuclear
interactions. Let us determine $\theta $ as an angle between tracks
of the primary and secondary particles. Using the PVD data for
different multiplicity intervals (small and high) which can be
considered as a certain value of the impact parameter, we have
discovered a two-hump structure. It was revealed in the extreme
\begin{figure}
\includegraphics[width=3.1 in, height=2.9 in, angle=0]{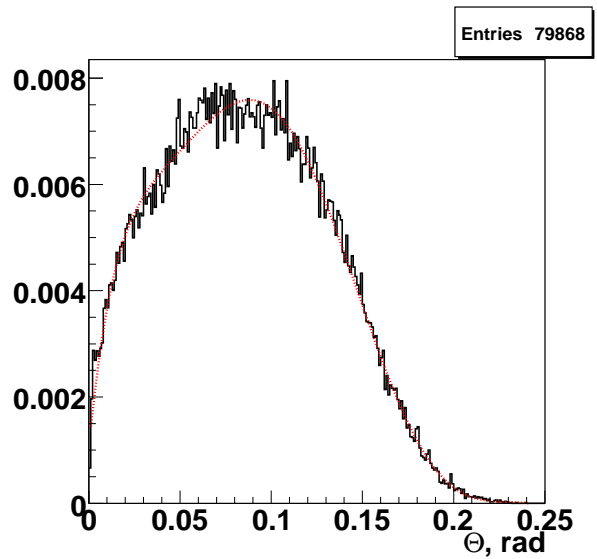}
\caption{\label{4}$\theta $ distribution for events with
multiplicity more than 8. The dotted curve is the approximation
obtained by using the seventh order polynomial.}
\end{figure}
multiplicity region. In Fig. 4 we give the $\theta $ distribution
for the events with more than 8 charged particles. The solid line is
the result of the polynomial approximation of the seventh order. We
see two different picks. These picks are absent in the case of small
multiplicity - not more than 8 charged particles (Fig. 5). In this
picture we compare these $\theta $ - distributions. The Monte-Carlo
simulation was carried out and high multiplicity events were
selected. The reconstruction of these events has shown the absence
of the two-hump structure for the $\theta $ - distributions (Fig.
6).
\begin{figure}
\includegraphics[width=3.1 in, height=2.9 in, angle=0]{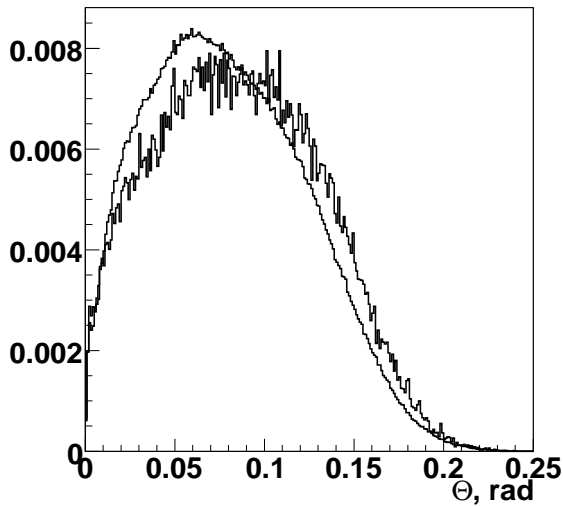}
\caption{\label{5}$\theta $ The $\theta $ distribution for events
with small, no more than 8 of charged particles (solid curve) and
with high multiplicity (the same two-hump curve like in Fig. 4).}
\end{figure}
\begin{figure}
\includegraphics[width=3.1 in, height=3.1 in, angle=0]{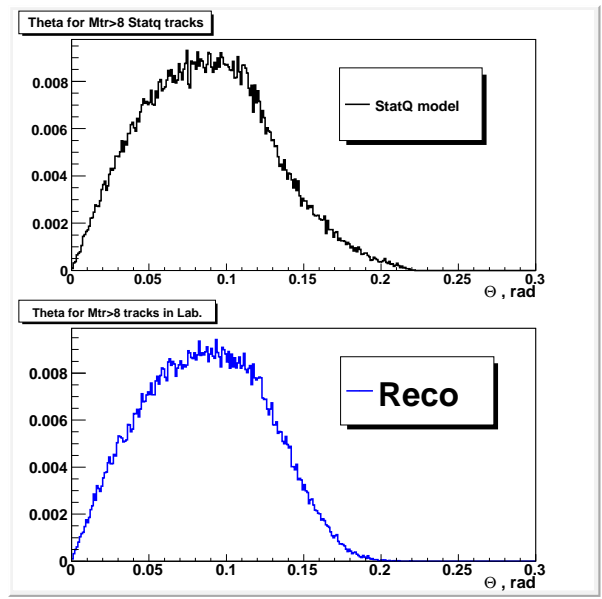}
\caption{\label{6}$\theta $ The $\theta $ distribution for
Monte-Carlo simulation for events with high multiplicity (more than
8): before (top) and after (bottom) reconstruction.}
\end{figure}
If we assume that this two-hump structure is caused by Cherenkov
gluon radiation, then it is possible to use the formula by  Dremin
\cite{Dremin} for the index of refraction. The value of $\theta _C $
determines angle $\theta $ between the direction one of the maximal
humps and the primary track: $\theta _C=0.065 \pm 0.005$ rad.
According to the Dremin theory for gluon rings \cite{Dremin},
\begin {equation}
cos \theta _C = 1/\beta n, \end{equation} where $\beta =p/\sqrt {p^2
+ m_p^2}$ and $n$ is the index of refraction. At beam momentum
$p$=50 GeV and proton mass $m_p$=0.938 GeV from (1) we obtain the
experimental value of the index of refraction at 50 GeV:
\begin {equation}
n= 1.0023 \pm 0.0003.
\end{equation}
Using the formula by Dremin we obtain the following:
$$
 n= 1+ \Delta n(p) =
$$
\begin {equation}
= \quad 1+ 3m_{pr}^2 \nu _h\sigma (p)\rho(p)/8\pi E_{pr},
\end{equation}
where $m_{pr}$ - mass of the parton, $\nu _h $ - the number of
scatterers within a single nucleon (conventional number is equal to
7), $\rho =Re F/Im F$ is the ratio of real and imaginal parts of the
scattering amplitude of partons, $E_{pr}$ - energy of the parton.
The parton mass and its momentum is replaced by the values for
proton: $m_{pr} \rightarrow m_p/\nu _h$ and $p_{pr} \rightarrow
p/\nu _h$. After that we get
\begin {equation}
\Delta n(p)=3m_p^3 Re F/2p^2= 0.0005 ReF.
\end{equation}
We can reach the agreement with our experimental result $\Delta
n=0.0023$ if $ReF$ = 4.6 $GeV^{-1}$ or 0.92 fm for parton scatters.

A possibility of of forming Bose-Einstein condensation (BEC)
formation in the extreme multiplicity region, has been shown in
\cite{Gorenstein} by Begun and Gorenstein. It is known that pions
(charged and neutral) are copiously formed at U-70 energies. They
are bosons. Their momenta are approaching to zero at high
multiplicity and the BEC can form. The pion number fluctuations will
be a prominent signal in the BEC-point. They predict that the scaled
variance of neutral and charged pion-number fluctuations $$\omega ^0
= <(\Delta N)^2>/<N>$$ in the vicinity of BEC-line, have an abrupt
and anomalous increase. Our project is aimed at checking this
prediction in the experiment.

Our Collaboration is preparing to check this prediction
experimentally. For this purpose we have selected the high
multiplicity events to determine the number of $\pi ^0$ (photon) for
every of them. Then the variance of the number of particle
fluctuations of the both neutral and charged pions may give a signal
about the BEC formation or its absence.

We are sure that the extreme multiplicity studies are very
interesting and useful. The recovering of the two-hump structure,
BEC formation, search for the turbulence and different collective
phenomena will be carried out at LHC and other centers.

We thank all participants of SVD Collaboration for active and
fruitful work on the project. The authors express deep recognition
to I.M. Dremin for stimulating discussion before these studies.
These investigations have been partially supported by RFBR grants
$08-02-90028-Bel\_a$ and $09-02-92424-KE\_a$.

\label{last}
\end{document}